\documentclass[aps,showpacs,twocolumn,amsmath,amssymb,pra]{revtex4-1} 

\usepackage{graphicx} 
\usepackage{bm}

\newcommand{\dd}[3]{\frac{\partial^{#3} #1}{\partial #2^{#3}}}
 
\begin{document} 
 
\title{Statistical analysis of complex systems with nonclassical invariant measures} 
 
\author{A. Fratalocchi$^{1,2}$} 

\email{andrea.fratalocchi@uniroma1.it} 
\homepage{www.primalight.org}

\affiliation{  
$^1$PRIMALIGHT, Faculty of Electrical Engineering; Applied Mathematics and Computational Science, King Abdullah University of Science and Technology (KAUST), Thuwal 23955-6900, Saudi Arabia\\
$^2$Dept. of Physics, Sapienza University of Rome, P.le A. Moro 2, 00185, Rome, Italy\\
}

\date{\today} 
 
\begin{abstract} 
I investigate the problem of finding a statistical description of a complex many-body system whose invariant measure cannot be constructed stemming from classical thermodynamics ensembles. By taking solitons as a reference system and by employing a general formalism based on the Ablowitz-Kaup-Newell-Segur scheme, I demonstrate how to build an invariant measure and, within a one dimensional phase space, how to develop a suitable thermodynamics. A detailed example is provided with a universal model of wave propagation, with reference to a transparent potential sustaining gray solitons. The system shows a rich thermodynamic scenario, with a free energy landscape supporting phase transitions and controllable emergent properties. I finally discuss the origin such behavior, trying to identify common denominators in the area of complex dynamics.
\end{abstract} 
 
 
\maketitle 

\section{Introduction}
The field of Statistical Mechanics (SM) delves with the problem of deducing macroscopic properties of matter stemming from microscopic states, specified by mechanical-like phase spaces whose dimensionality depends on the atomic degrees of freedom of the system. Classical SM lies its foundations in the orthodicity problem posed by Boltzmann in the XIX century \cite{bo68}, and in particular in the search of \emph{invariant measures} \footnote{An invariant measure is a probability distribution, defined in a suitable phase space, which behaves as a conserved quantity during the time evolution of the system.} whose infinitesimally change leads, in the thermodynamic limit, to thermodynamic relations among internal energy, volume, pressure and average kinetic energy. Boltzmann's work leads to three models of thermodynamics, namely the microcanonical, canonical and grand canonical ensembles, each characterized by a diverse invariant measure of the phase space \cite{SMAST}. In more recent times, the same invariant measures have been found by Shannon in a famous paper \cite{sha0}, by employing a rather different approach based on an variational minimization problem connected to an information functional defined axiomatically. Boltzmann and Shannon's approach has in common the search of invariant measures under specific constraints (i.e., the thermodynamic relations or the minimization of the information functional), whose role is to define models of thermodynamic with the minimum information possible on the microscopic state. However, a natural question therefore arise, and concerns the possibility to develop a SM description of microscopic mechanical systems whose invariant measure is not constrained by any a-priori condition. This problem acquires fundamental importance in the field of Complex Systems (CS). These are characterized by the presence of a large number of interconnected or interwoven parts, which act together to sustain emergent properties of the whole ensemble \cite{ybaym}. Illustrative examples of CS are found in many different fields including ---and not limited to--- neural networks, colloids, complex liquids and glasses \cite{SGTAB,ybaym,complx,RFASG,SGTAB,ybaym,goetze91:_liquid_freez_and_glass_trans}. Contrary to classical thermodynamic systems, mainly described within two scales (micro and macro), complex systems exhibit a full set of intermediate length scales each characterized by specific properties (see e.g. \cite{monaco09:_anomal_proper_of_acous_excit} in the context of structural glasses). In this article I focus my analysis on the so called mesoscopic scale \cite{ybaym}, which contains a large (but finite) number of atomic parts $n$. This length scale allows for a general treatment of the problem as it does not require any hypothesis on the structure of the invariant measure at the thermodynamic limit, when $n\rightarrow\infty$. With these premises, my initial question becomes: \emph{with reference to a dynamical many-body ensemble, whose invariant measure is not constrained by any a-priori condition, could a statistical mechanic framework be developed at the mesoscopic scale?}\\
An important benchmark for such problem is provided by the theory of solitons, which are exact, localized particle-like solution of nonlinear systems integrable through the inverse scattering transform \cite{SATIST,DACNSS}. Although there are well documented observations of soliton waves in the past centuries \cite{rus0}, their mathematical description has become clear only in recent times \cite{kdv_is_gardea}. Owing to their robust particle nature, solitons are preferential carrier of energy transport and are therefore ubiquitous in physics \cite{WII,SICMP,GS,SIMAP,OSFFTPC,KS75,PhysRevLett.84.3740,PhysRevLett.84.1070,bf_solg_ela,fratalocchi:044101}. 
According to our initial question, we ask what is the invariant measure of a mesoscopic many-body ensemble of solitons, its relationship with canonical ensembles and, more important, how to develop a proper statistical mechanics framework to describe the appearance of (any) emergent dynamics.
The development of such theory is not only of interest to the field of mathematical physics and statistical mechanics, but could be the backbone for novel studies aimed at describing extreme events occurring in nonlinear dynamics, such as e.g., the formation of rogue waves, which are attracting a large scientific interest \cite{rg0,org0}.\\
This paper is organized as follows. Section \ref{akns0} presents a general formulation of the problem within the Ablowitz-Kaup-Newell-Segur scheme. An invariant measure of the soliton system is calculated in Sec. \ref{ssiv0}, while its thermodynamics analysis is developed in Sec. \ref{ssth0}. The general theoretical framework is then applied to a specific case in Sec. \ref{ist}, encompassing both a thermodynamic and time-domain analysis. Sec. \ref{sec:cm}, finally, discusses the origin of the solitons complex behavior, highlighting similarities and differences with other systems showing complex dynamics.

\section{General Formulation through the AKNS scheme}
\label{akns0}
In order to pursue a general theory, I employ the Ablowitz-Kaup-Newell-Segur (AKNS) scheme \cite{AKNS0} whose general framework encompasses a large selection of integrable equations in both $(1+1)$ and $(2+1)$ dimensions, including the Korteweg-de Vries (KdV), the Kadomtsev-Petiashvili (KP), the Sine-Gordon (SG) and Nonlinear Schr\"odinger (NLS) equations. I begin by introducing the pair of equations:
\begin{align}
 \label{xxx}
  &\bigg[\frac{\partial}{\partial x}-U\bigg]F=0,\\
\label{ttt}
&\bigg[\frac{\partial}{\partial t}-A\bigg]F=0,
\end{align}
with $F=\begin{bmatrix}&\psi_1\\&\psi_2\end{bmatrix}$ a two-element vector and the $2\times 2$ matrices:
\begin{align}
&  U=\begin{bmatrix}
&-i\lambda &q\\
&r &i\lambda
\end{bmatrix},\nonumber\\
& A=\begin{bmatrix}
&A_{11} &A_{12}\\
&A_{21} &A_{22}
\end{bmatrix},
\end{align}
being $\lambda$ the spectral parameter, $q(x,t)$, $r(x,t)$ generic complex potentials and $A_{ij}=A_{ij}(x,t,\lambda)$ unknown functions of the spectral parameter $\lambda$. The compatibility between Eqs. (\ref{xxx}) and (\ref{ttt}) yields the zero curvature condition:
\begin{align}
\label{zcc0}
  \frac{\partial}{\partial x}A-\frac{\partial}{\partial t}U+[A,U]=0,
\end{align}
which gives an infinite hierarchy of integrable nonlinear partial differential equations generally expressed as $f(q,r)=0$. To ensure the existence of $F$ on the infinite line $x\in [-\infty,\infty]$, I need to specify the behavior of the potentials $q$ and $r$ at the boundaries $\lvert x\rvert\rightarrow\infty$. Following the original paper of Ablowitz \emph{et al.} \cite{AKNS0}, and without loss of generality, I assume that both $q$ and $r$ decay sufficiently rapidly to $0$ (we will see in Sec. \ref{ist} how to handle the finite density case, i.e., $q,r\rightarrow\rho e^{i\phi}$ for $x\rightarrow\pm\infty$). 

\subsection{Looking for an invariant measure}
\label{ssiv0}
The solution $F$ can be expressed by a linear combination of the Jost functions $F_\pm(x,\lambda)$, whose asymptotic behavior is found by solving (\ref{xxx}) for $\lvert x\rvert\rightarrow\infty$:
\begin{align}
  &F_+(+\infty,\lambda)=\sigma_0 e^{-i\lambda\sigma_3 x}, &F_-(-\infty,\lambda)=\sigma_3 e^{-i\lambda\sigma_3 x},
\end{align}
being $\sigma_i$ ($i=1,3$) one of the $SU(2)$ generators ($\sigma_0$ is the identity matrix):
\begin{align}
  &\sigma_0=\begin{bmatrix}
1 &0\\
0 &1
\end{bmatrix}, &
\sigma_1=\begin{bmatrix}
0 &1\\
1 &0
\end{bmatrix},\nonumber\\
  &\sigma_2=\begin{bmatrix}
0 &-i\\
i &0
\end{bmatrix}, &
\sigma_3=\begin{bmatrix}
1 &0\\
0 &-1
\end{bmatrix}.
\end{align}
Each Jost function $F_\pm$ provides a basis for the solution of Eqs. (\ref{xxx}), hence, they are linearly dependent and connected via the scattering matrix $T(\lambda)$:
\begin{align}
\label{tl0}
&  F_-=F_+T(\lambda), &T=\begin{bmatrix}&a(\lambda) &b^*(\lambda)\\&b(\lambda) &-a^*(\lambda)\end{bmatrix}.
\end{align}
The knowledge of $T$ yields an equivalent representation of the dynamics $f(q,r)=0$ via the spectral transform $\mathcal{S}$:
\begin{align}
\label{stt0}
  \mathcal{S}(q,r)=\begin{cases}
    R(\lambda)=\frac{b(\lambda)}{a(\lambda)},     \\
    \lambda_n, \\
    c_n=b(\lambda_n)\bigg[\frac{d a(\lambda_n)}{d\lambda}\bigg]^{-1},
    \end{cases}
\end{align}
with $-\infty<\lambda<\infty$ and $n=0,...,N$. The transform $\mathcal{S}$ is composed of:\\
i) the reflection coefficient $R$, which composes the continuum spectrum describing radiation;\\
ii) the discrete eigenvalues $\lambda_n$ [originated from the zeros of $a(\lambda)$] and the normalization constants $c_n$. This constitutes the discrete spectrum defining solitons.\\
Thanks to the spectral transform, the dynamics of both $q$ and $r$ is expressed in terms of the time evolution of the spectral coefficients $a$ and $b$. From the zero curvature condition (\ref{zcc0}), straightforward calculations \cite{AKNS0,SAI} yield:
\begin{align}
\label{abev}
&  a(\lambda,t)=a(\lambda,0)=a(\lambda), &b(\lambda,t)=b(\lambda,0)e^{-2\alpha t}
\end{align}
being $\alpha(\lambda)$ a generic function whose form depends on the specific equation considered. According to Eqs. (\ref{abev}), the scattering coefficient $a(\lambda)$ is a constant of motion and can be therefore employed to derived an invariant measure for the overall system. This can be obtained from the soliton discrete eigenvalues $\lambda_i$, which constitute the time-invariant part of the spectral transform (\ref{stt0}). 
More specifically, considering the one dimensional phase space given by the infinite line $\lambda\in[-\infty,\infty]$, I define the soliton invariant measure $\mu(\lambda)$ from the eigenvalues Density Of States (DOS):
\begin{align}
\label{solm0}
  \mu(\lambda)=\frac{1}{N}\sum_{j=1,N}\delta(\lambda-\lambda_j)
\end{align}
with $\lambda_j$ being a soliton eigenvalue. The advantages of (\ref{solm0}) are twofold:\\
i) it yields an exact description of the dynamics of soliton waves, as they originate from the discrete spectrum which depends on $\lambda_n$;\\
ii) it is defined on a single dimensional phase space. This allows for a simpler analysis with respect to the time dependent description of $q$, $r$, which gives rise to infinite dimensional phase spaces harder to handle.\\
It is important to stress that the invariant measure (\ref{solm0}) exists only in nonlinear systems supporting soliton waves. In linear regime, in fact, the spectral transform reduces to the reflection coefficient $R$ only \cite{calogero82:_spect_trans_and_solit}. This quantity is not an invariant of motion (see Eqs. \ref{abev}), and no spectral invariant measure can be built in this case.\\
In summary, a probability function based on the conserved part of the spectral transform is an invariant measure for an ensemble of solitons, even in the presence of radiation, which allows for the development of an exact system description within a one dimensional phase space. 
  
\subsection{Mesoscale thermodynamics through Lie transformation groups}
\label{ssth0}
Equations (\ref{stt0})-(\ref{solm0}) possess the following two properties:\\
i) Solitons eigenvalues $\lambda_n$ are arbitrary in the interval $\lambda\in [-\infty,+\infty]$. The inverse scattering, in fact, guarantees a $1:1$ relationship between a chosen distribution of soliton DOS and a scattering potential $\psi_0$, which may be calculated by solving the corresponding Gelfand-Levitan-Marchenko integral equations \cite{SATIST}. In consequence of that, the soliton invariant measure (\ref{solm0}) does not satisfy any entropy maximization principle and could not be expressed in terms of the three classical thermodynamic ensembles;\\
ii) Any invariant of motion, including the Hamiltonian, follows the single-soliton factorization:
\begin{align}
\label{nint0}
&  \mathcal{H}=\sum_n\mathcal{H}_n, &\mathcal{H}_n=\mathcal{H}_n(\lambda_n),
\end{align}
which stems from the factorization property of the spectral transform:
\begin{align}
&  \mathcal{S}=\{n=0,...,N,\lambda_n,c_n\}=\sum_{n=0}^N\mathcal{S}_n, &\mathcal{S}_n=\{\lambda_n,c_n\},
\end{align}
and yields a microscopic mechanical system (a soliton gas) described by a purely kinetic Hamiltonian:
\begin{equation}
  \label{eq:pp}
  \mathcal{H}=\sum_n\mathcal{H}_n(\lambda_n)\equiv\sum_n\frac{p_n^2}{2}.
\end{equation}
 It is worth to observe that the wave-particle ``dualism'' originated by Eqs. (\ref{zcc0}) and (\ref{eq:pp}) ---i.e., a wave field $[q(x,t),r(x,t)]$ associated to particle modes $\lambda_j$ with momenta $p_j$--- goes beyond the ordinary wave-particle dualism of classical optics or quantum mechanics, and is only manifested in nonlinear equations exhibiting soliton waves. In quantum mechanics, for example, due to the Heisenberg uncertainty principle it would never be possible to observe a double localization ---i.e., the appearance of position (momentum) eigenmodes which are localized in momentum (position)--- or localized wavepackets able to overwhelm spreading effects. For soliton waves, conversely, we do observe an ensemble of kinetic eigenmodes with specific momenta $p_n=\sqrt{2\mathcal{H}_n}$, whose spatial form remains localized in space. This is a peculiar property of solitons and stems from their pure nonlinear nature.\\
In order to provide a statistical mechanics analysis of a solitons ensemble, we come back to our original question of how to describe of a microscopic mechanical system under a non canonical measure. I address this problem by mutuating ideas from the thermodynamics of chaos \cite{TOCS}. In classical statistical mechanics a special role is deserved to canonical probability distributions, as they form the only ensemble (canonical) that satisfies the thermodynamics relations even outside the thermodynamic limit. Another important property of canonical distributions, which I will exploit here, is their algebraic structure. I begin by considering the soliton DOS (\ref{solm0}), defined on the line interval $\lambda\in[\lambda_{min},\lambda_{max}]$, and partition the latter into $N_\epsilon=(\lambda_{max}-\lambda_{min})/\epsilon$ boxes of size $\epsilon$. I then define $p_i$ as the probability of finding an eigenvalue in the $i-$th box:
\begin{align}
&  p_i=\int_{\lambda_{min}+(i-1)\epsilon}^{\lambda_{min}+i\epsilon}\mu(\lambda)\mathrm{d\lambda}, &i=1,2,...,N_\epsilon.
\end{align}
Then, the application of a canonical measure on $p_i$ [a special case of Escort Distributions (ED) \cite{TOCS}]:
\begin{align}
\label{esc0}
  P_i=f(p_i,\beta)=\frac{p_i^\beta}{\sum_j  p_j^\beta},
\end{align}
yields a Lie group \cite{SAIMFDE,AOLGTDE} of transformations acting on the space $p_i$, with the infinite dimensional parameter of transformation $\beta=1/T$ playing the role of an inverse temperature. The identity element is obtained for $\beta=1$, while straightforward calculations yield the Lie group composition law function $\phi$, which reads $\phi(\beta_1,\beta_2)=\beta_1\beta_2$. Such a Lie group of transformations can be linked to the field of statistical mechanics by defining the following partition function:
\begin{align}
\label{zzz}
  Z(\beta)=\sum_j p_j^\beta= \exp(-\Psi).
\end{align}
Thanks to this definition, in fact, the representative probability $P_i$ becomes:
\begin{align}
 P_i=\exp(\Psi-\beta \mathcal{E}_i), 
\end{align}
with  $\mathcal{E}_i=-\ln p_i$ acting as the energy of the $i$-th microstate. The Helmholtz free energy $\mathcal{F}$ is then given by the corresponding expression in the canonical ensemble:
\begin{align}
\label{fff}
  \mathcal{F}(\beta)=\Psi/\beta.
\end{align}
To develop a mesoscale thermodynamics, I study how the Free-energy approaches the thermodynamic limit, defined by 
\begin{equation}
\label{fth0}
\beta \mathcal{F}=\mathcal{X}=\lim_{V\rightarrow\infty}\frac{\Psi}{V}
\end{equation}
with $V=-\ln\epsilon$ being the volume. For $\epsilon\rightarrow 0$, the corresponding partition function scales as:
\begin{align}
  \label{fth1}
  &Z\sim\epsilon^\chi.
\end{align}
It is worth to observe that the calculation of the rescaled Free Energy $\chi$ does not need the real computation of the thermodynamic limit $V\rightarrow\infty$, but is obtained from the scaling of the partition function as the volume increases.\\
In conclusion, in the case of a mechanical phase space possessing an arbitrary measure, a statistical description is possible by exploiting the Lie group structure of the canonical ensemble. This yields a general mesoscale thermodynamics framework [i.e., Eqs. (\ref{fff})-(\ref{fth1})]. In order to illustrate these concepts, I will consider in the next section a specific example.

\section{Application to the Nonlinear Schr\"odinger Equation}
\label{ist}
Among the various integrable equations described by the AKNS scheme, I discuss a universal model of nonlinear wave propagation in dispersive media, namely the NLS equation:
\begin{align}
\label{nlsd0}
  i\frac{\partial\psi}{\partial t}-\frac{\partial^2\psi}{\partial x^2}+2(\lvert\psi\rvert^2-\rho^2)\psi=0,
\end{align}
on the infinite line $x\in[-\infty,\infty]$ with the general boundary condition:
\begin{align}
\label{boo1}
  &\psi\rightarrow\rho e^{i\phi}, &x\rightarrow\pm\infty.
\end{align}
The NLS arises from the AKNS system by taking:
\begin{align}
  U&=\frac{1}{2}\big[\sigma_1\big(\psi^*+\psi\big)+i\sigma_2\big(\psi^*-\psi\big)\big]+\lambda\frac{\sigma_3}{2i},\nonumber\\
A&=i\sigma_3\big(\lvert\psi\rvert^2-\rho^2\big)-\frac{1}{2}\nonumber\\
&\times\bigg[i\sigma_1\bigg(\dd{\psi^*}{x}{}+\dd{\psi}{x}{}\bigg)-\sigma_2 \bigg(\dd{\psi^*}{x}{}-\dd{\psi}{x}{}\bigg)\bigg]-\lambda U.
\end{align}
The universality of this model \cite{WII} makes it extremely interesting on the experimental side \cite{BEC,Kivshar98,OSFFTPC,sw_so_ghofra,fleisher07,sw_ds_hoefe,PhysRevA.69.043610,PhysRevA.69.063605,el:053813}, while its specific nonlinear sign (defocusing) allows to extend the application domain of this thermodynamics framework to the general boundary conditions expressed by (\ref{boo1}).

\subsection{Spectral transform of a reflectionless scattering potential}
\label{sp}
A reflectionless potential is, by definition, an input potential $\psi_0=\psi(x,0)$ whose spectrum contains just the discrete part (i.e., soliton waves). For the defocusing NLS, I consider the input:
\begin{align}
  \label{gray1}
  &\psi_0=-\rho(w+iv)[w\tanh(wx)+iv], &w^2+v^2=1.
\end{align}
When $\rho=1$, Eq. (\ref{gray1}) yields the one-soliton solution of the NLS with ''grayness'' $w$ and velocity $2v$ \cite{HMITTOS,kivshar98:_dark_optic_solit}. In the following, I will consider the general case $\rho\neq 1$ and find the conditions to have a reflectionless potential. I begin my analysis with the spectral problem:
\begin{align}
  \dd{F}{x}{}=U(\psi_0,x,\lambda)F,
\end{align}
also known as linear auxiliary problem \cite{HMITTOS}, and in particular by finding its Jost solutions $F_\pm$. These are the solutions of the following system:
\begin{align}
  \label{lap0}
&\frac{\partial F_\pm }{\partial x}=\bigg(\psi_0\sigma_1+\frac{\lambda}{2i}\sigma_3\bigg)F_\pm , &\psi_0\in\Re,
\end{align}
being $F_\pm(x,\lambda)$ a $2\times 2$ matrix with asymptotic conditions $F_\pm (\pm \infty,\lambda)\sim F_{\pm \infty}$, the latter calculated from (\ref{lap0}) at $x\rightarrow\pm\infty$:
\begin{align}
&F_{-\infty}=\begin{bmatrix}
  1 &\frac{\omega}{i(\lambda+k)}\\
\frac{i\omega}{(\lambda+k)} &1
\end{bmatrix}
e^{-\frac{ikx}{2}\sigma_3},\nonumber\\
&F_{+\infty}=e^{-i\frac{\theta}{2}\sigma_3}F_{-\infty},
\end{align}
being $k=\sqrt{\lambda^2-4\rho^2}$. 

\subsubsection{Case $v=0$}
\label{sec:dl0}
When $v=0$, the input $\psi_0$ turns into the so called tanh-like potential, which is known to be integrable \cite{kivshar98:_dark_optic_solit,Zhao:89,Zhao:89b}. Nevertheless, a complete derivation of the spectral transform is missing in the literature. For this reason, I will provide a thorough analysis of this limiting condition in the following paragraphs. This derivation will be also helpful when considering the more general case $v\neq 0$.\\
I begin by calculating the Jost solution $F_-$. I apply the transformation:
\begin{align}
\label{tra0}
&F_-(x,\lambda)=F_{-\infty}E(x,\lambda),
\end{align}
with $E$ being a $2\times 2$ matrix of components:
\begin{align}
  E=\begin{bmatrix}
E_{11} &E_{12}\\
E_{21} &E_{22}
\end{bmatrix}
\end{align}
and asymptotic value
\begin{align}
 E(-\infty,\lambda)=\begin{bmatrix}1&0\\0&1\end{bmatrix}=\sigma_0, 
\end{align}
thus obtaining a linear auxiliary problem for $E$:
\begin{align}
\label{sp0}
  \frac{\partial^2 E}{\partial x^2}=\frac{\psi_0-\rho}{k}\big[2i\rho\sigma_3+\lambda\sigma_1e^{-ikx\sigma_3}\big]E.
\end{align}
By employing the change of coordinate
\begin{align}
 s=\frac{1+\tanh x}{2}, 
\end{align}
 I transform (\ref{sp0}) into an Hypergeometric equation for $E_{11}$:
\begin{equation}
  s(1-s)\frac{\partial^2 E_{11}}{\partial s^2}-\bigg(s+i\frac{k}{2}\bigg)\frac{\partial E_{11}}{\partial s}+\rho^2 E_{11}=0,
\end{equation}
defined by the following Riemann $\mathcal{P}$ symbol:
\begin{align}
\label{hyp0}
  E_{11}=\mathcal{P}\begin{Bmatrix}
0 &1 &\infty\\
0 &0 &-\rho\\
1+i\frac{k}{2} &-i\frac{k}{2} &\rho
\end{Bmatrix}.
\end{align}
The symmetry properties of (\ref{sp0}):
\begin{equation}
  E_{21}=E_{11}|_{k\rightarrow -k},
\end{equation}
 and:
 \begin{align}
   &E_{22}=E_{11}^*, &E_{12}=E_{21}^*,
 \end{align}
allows to express all the elements of the unknown matrix $E$ through the single term $E_{11}$ defined by (\ref{hyp0}). In particular, from the properties of Hypergeometric functions near $s=0$ \cite{ODE} (i.e., for $x\rightarrow -\infty$):
\begin{widetext}
\begin{align}
  E=\begin{bmatrix}
&_2F_1\bigg(-\rho,\rho,i\frac{k}{2},s\bigg) &_2F_1\bigg(1-\rho+i\frac{k}{2},1+\rho+i\frac{k}{2},2+i\frac{k}{2},s\bigg)s^{1+i\frac{k}{2}}\cdot c&\\
&_2F_1\bigg(1-\rho-i\frac{k}{2},1+\rho-i\frac{k}{2},2-i\frac{k}{2},s\bigg)s^{1-i\frac{k}{2}}\cdot c &_2F_1\bigg(-\rho,\rho,-i\frac{k}{2},s\bigg)
\end{bmatrix}
\end{align}
\end{widetext}
As the reader can verify by straightforward algebra, in the limit $s\rightarrow 0$, $E\rightarrow \sigma_0$. The constant $c$ is to be determined from the prolongation of $F_-$ to $x\rightarrow +\infty$, needed to relate $F_-$ to $F_+$.  As stated in (\ref{tl0}), the Jost solutions $F_\pm$ are not independent and are related through the linear scattering matrix $T(\lambda)$. The latter, considering our boundary conditions, is expressed as \cite{HMITTOS}:
\begin{align}
\label{mpm0}
&F_-(x,\lambda)=F_+(x,\lambda)T_\rho(\lambda), \nonumber\\
&T_\rho(\lambda)=\begin{bmatrix}
a(\lambda) &b^*(\lambda)\\
b(\lambda) &a^*(\lambda)
\end{bmatrix},
\end{align}
with the transition coefficients $a(\lambda)$ and $b(\lambda)$ related by $\lvert a\rvert^2-\lvert b\rvert^2=1$.  From the latter condition and through the analytical continuation of the Hypergeometric functions near $s=1$ ($x\rightarrow +\infty$), I obtain the the constant $c=2i\lambda\rho/k(2i+k)$ and the transition coefficients:
\begin{align}
\label{rc0}
  &a=i\frac{k\cdot\Gamma(\frac{k}{2i})^2}{\lambda\cdot\Gamma(\frac{k}{2i}+\rho)\Gamma(\frac{k}{2i}-\rho)},\nonumber\\
  &b=\frac{k\cdot\lvert\Gamma(i\frac{k}{2})\rvert^2}{2\Gamma(1-\rho)\Gamma(\rho)}.
\end{align}
The coefficients $a$ and $b$ of the scattering matrix allow to calculate the spectral transform $\mathcal{S}\{\psi_0\}$ of the real input pulse $\psi_0$. In particular, for integer values of $\rho$:
\begin{align}
  &R(\lambda)=0, \nonumber\\
&\lambda_{\pm n}=\pm 2\sqrt{\rho^2-(\rho-n)^2}, &n=0,\dotsc,M
\end{align}
while $c_n$ can be calculated from \cite{HMITTOS}:
\begin{align}
c_n=\frac{b(\lambda_n)}{\dot{a}\bigg(\lambda_n+i\sqrt{4\rho^2-\lambda_n^2}\bigg)}.  
\end{align}
and reads (see App. \ref{cn0}):
\begin{align}
\label{cn1}
  &c_{\pm n}=(-1)^{n+1}\frac{\prod_j(s_{\pm n}-s_j^*)}{\prod_{j\neq \pm n}(s_{\pm n}-s_j)}, &n=0,\dotsc,M
\end{align}
with $s_{\pm n}=\lambda_{\pm n}+2i(\rho-n)$, $j\in[-M,M]$ and $M=\rho-1$. The spectral transform $\mathcal{S}\{\psi_0 \}$, for integer $\rho$, is therefore expressed as:
\begin{align}
\label{stt}
&\mathcal{S}\{\psi_0\}=\begin{cases}
R(\lambda)=0,\\
\lambda_{\pm n}=\pm 2\sqrt{\rho^2-(\rho-n)^2},\\
c_{\pm n}=(-1)^{n+1}\frac{\prod_j(s_{\pm n}-s_j^*)}{\prod_{j\neq \pm n}(s_{\pm n}-s_j)},
\end{cases}
\end{align}
with $n=0,\dotsc,\rho-1$.
\begin{figure}
\includegraphics[width=8.5 cm]{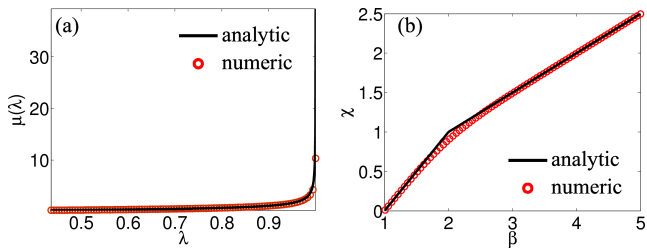}
\caption{\label{pv0} (Color Online). (a) Theoretical (solid line) and numerical (red circles) soliton invariant measure versus $\lambda/2\rho$ calculated for grayness $w=0.9$; (b) Theoretical (solid line) and numerical (red circles) free energy $\mathcal{X}$ versus inverse temperature $\beta$ for $w=0.9$.}
\end{figure}
\subsubsection{General case $v\neq 0$}
By applying the coordinate change $s=(1+\tanh wx)/2$, with the F-Homotopic transformation $(E_-)_{11}=(s-1)^{-i\frac{k}{4w}}s^{-i\frac{k}{4w}}\varphi$, I reduce Eq. (\ref{lap0}) to the normal form of Heun equation \cite{HDE}:
\begin{align}
\label{psy}
  \varphi=\mathcal{P}\begin{Bmatrix}
&0 &1 &c &\infty &\\
&0 &0 &0 &\alpha &s &q\\
&1-\gamma &1-\delta &2 &\beta &\\
\end{Bmatrix},
\end{align}
with $c=(w-iv)/2w$, $\gamma=\delta= 1-i k/2 w$, $\beta=-i k/2 w+\rho$, $\alpha=-i k/2 w-\rho$, and $q$ being the \emph{accessory} parameter whose cumbersome expression will not be reported here. The Jost solutions and the $a$ and $b$ coefficient of the scattering matrix can be then constructed by applying the same analysis of Sec. \ref{sec:dl0}. However, the \emph{connection problem} associated with the Heun equation, i.e. the analytic prolongation of Heun functions, is to date unsolved \cite{HDE}. A complete derivation on the spectral transform is then not possible in this case, due to the impossibility of calculating the zeros of $a$ and $b$ in analytic form. However, closed form expression for the radiation spectrum and the soliton eigenvalues are accessible by exploiting an original expansion with Hypergeometric functions \cite{a.:_time_rever_refoc_of_expan}. In particular, for integer $\rho$:
\begin{align}
\label{eigda}
S\{\psi_0\}=\begin{cases}
  R(\lambda)=0,\\
  \lambda_{\pm n}=\pm 2\sqrt{\rho^2-(\rho-n)^2w^2},\\
  \lambda_0=2v\rho,
\end{cases}
\end{align}
with $n=1,...,\rho-1$.\\
The scattering potential $\psi_0$, expressed by Eq. (\ref{gray1}), is therefore a reflectionless potential for integer amplitudes $\rho$, and contains $2\rho-1$ soliton particles whose eigenvalues are expressed by (\ref{eigda}).

\subsection{Mesoscale thermodynamics}
\label{sec:th0}
The soliton invariant measure, for the scattering potential $\psi_0$, reads:
\begin{align}
  \label{dens0}
  \mu(\lambda)=\frac{1}{2\rho-1}\sum_{m=1-\rho}^{\rho-1}\delta(\lambda-\lambda_m)
\end{align}
having ordered the eigenvalues (\ref{eigda}) in ascending order. At the mesoscopic scale $\rho\gg 1$, I further simplify the soliton DOS by introducing the variable $y=(1-n/\rho)w$, continuous in the range $[0,w]$, and then rescale the soliton eigenvalues $\lambda\rightarrow\lambda\cdot2\rho$, with $|\lambda|\in[\sqrt{1-w^2},1]$. With this position, for $\lambda>0$, Eq. (\ref{dens0}) becomes:
\begin{align}
  \label{dens1}
  \mu(\lambda)=\frac{\lambda}{2w\sqrt{1-\lambda^2}}.
\end{align}
This analytical expression compares well with numerical estimates (Fig. \ref{pv0}-a). As seen from Eq. (\ref{dens1}), the soliton invariant measure depends on two parameters: the soliton eigenvalue $\lambda$ and the grayness $w$; the latter condition adds one more coordinate to the free energy $\chi=\chi(\beta,w)$, which becomes function of two different variables.
\begin{figure}
\includegraphics[width=8.5 cm]{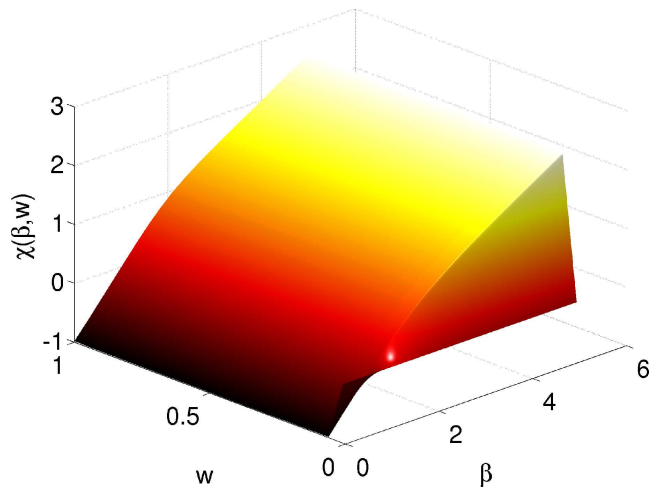}
\caption{\label{fel} (Color Online). Free energy profile for varying temperature $1/\beta$ and grayness $w$.}
\end{figure}
\subsubsection{Free energy $\chi$ versus temperature: Emergence of phase transitions}
I begin by studying the structure of the free energy landscape for a varying temperature $1/\beta$, thus analyzing the phase transition scenario of the system. In order to apply the thermodynamic framework discussed in Sec. \ref{ssth0}, I first partition the spectral gap $[-1,1]$ into $N_\epsilon=2/\epsilon$ boxes of length $\epsilon$ and then calculate the probability $p_i$ of finding an eigenvalue in the box $[-1+(i-1)\epsilon,-1+i\epsilon]$:
\begin{align}
\label{pj}
  p_i=\int_{(i-1)\epsilon}^{i\epsilon}&\mu(\lambda)\mathrm{d\lambda}=\frac{2\rho}{2\rho-1}\sum_{n=1-\rho}^{\rho-1}\nonumber\\
&\times\int_{-1+(i-1)\epsilon}^{-1+i\epsilon}\mathrm{d\lambda}\delta(\lambda-\lambda_n)\propto\nonumber\\
&\sqrt{1-(i-1)^2\epsilon^2}-\sqrt{1-i^2\epsilon^2},
\end{align}
where in the last step Eq. (\ref{dens1}) has been used. When $\epsilon\rightarrow 0$, two principal scaling regimes in (\ref{pj}) emerge. In particular, in the case of $|\lambda|\approx 1$, an expansion of $p_i$ for $i\approx 0$ yields at leading order a factor proportional to $\sqrt\epsilon$, while for $|\lambda|\approx 0.5$ the leading order of the expansion becomes of $O(\epsilon)$. In summary:
\begin{align}
  \begin{cases}
p_j\propto\epsilon \;\;\;\text{  for  } \;|\lambda|\approx0.5, \nonumber\\
p_j\propto\sqrt{\epsilon} \;\;\;\text{  for  } \;|\lambda|\approx 1.
\end{cases}
\end{align}
The partition function $Z$ then scales as:
\begin{align}
Z=\sum_j p_j^\beta\sim (N_\epsilon-2)\epsilon^{\beta}+&2\epsilon^{\beta/2},
\end{align}
with two boxes near $\lambda=\pm 1$ scaling like $\epsilon^{\beta/2}$, and the remaining $N_\epsilon -2$ scaling like $\epsilon^{\beta}$. In the limit $\epsilon\rightarrow 0$, the partition function reads:
\begin{align}
  Z\sim\epsilon^{\beta-1}+\epsilon^{\beta/2}\sim \epsilon^\mathcal{X}.
\end{align}
This implies $\mathcal{X}=\mathrm{min}[\beta-1,\beta/2]$ and leads to:
\begin{align}
  \label{free0}
\mathcal{X}=\begin{cases}
\beta-1, &\beta\leq \beta_c\\
\frac{\beta}{2}, &\beta>\beta_c
\end{cases}
\end{align}
being $\beta_c=2$ a critical     point      given      by
$\beta_c-1=\beta_c/2$. The free energy (\ref{free0}), being continuous but not differentiable at $\beta_c$, supports a  first order phase  transition as the  inverse temperature $\beta$ decreases below $\beta_c=2$. As  seen  in Fig.  (\ref{pv0}-b),  this result is in agreement with the numerically calculated free energy $\mathcal{X}$ from the soliton invariant measure (\ref{dens0}). 
\begin{figure}
\includegraphics[width=8.5 cm]{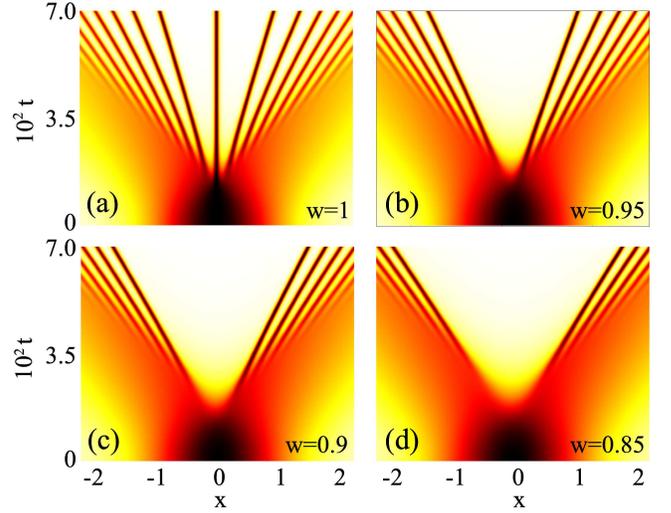}
\caption{\label{ss1} (Color Online). Time dependent dynamics of the many-body soliton ensemble for $w=1$ (a), $w=0.95$ (b), $w=0.9$ (c), $w=0.85$ (d). In all simulations the amplitude of the input $\psi_0$ is set to $\rho=30$.}
\end{figure}

\subsubsection{The complete scenario: controllable emergent properties and free-energy collapse.}
\begin{figure}
\includegraphics[width=7 cm]{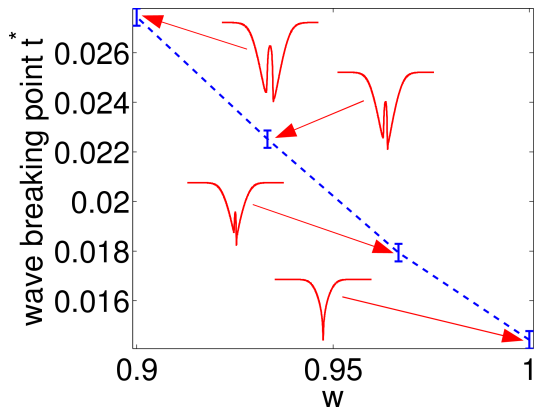}
\caption{\label{ss2} (Color Online). Time position of the wave breaking point (dashed line) versus grayness $w$, with insets showing the corresponding field intensity profiles (solid lines).}
\end{figure}

The effects of the grayness parameter $w$ on the dynamics are then evaluated by inspecting the soliton phase space structure. In particular, a decreasing of $w$ from $1$ to $0$ is accompanied by the shrinking of the the phase space region $|\lambda|\in[\sqrt{1-w^2},1]$  containing solitons, with the consequent compression of soliton eigenvalues on spectral gap edges at $\lambda=\pm 1$. Such process leads to the collapse of the invariant measure towards two eigenvalues:
\begin{align}
  \lim_{w\rightarrow 0}\mu(\lambda)\approx \frac{1}{2}[\delta(\lambda-1)+\delta(\lambda+1)].
\end{align}
In this case the corresponding partition function scales as:
\begin{align}
  \lim_{w\rightarrow 0}Z=\sum_j^{N_\epsilon} p_j^\beta=2\left(\frac{1}{2}\right)^\beta\sim\sum_j\epsilon^0\sim\epsilon^{\mathcal{X}(\beta)},
\end{align}
which predicts the emergence of a null free energy $\chi=0$. The net effect of decreasing $w$ is therefore to increase the ''pressure'' of soliton eigenvalues towards spectral gap edges, thereby leading to the collapse of the whole free energy landscape to a flat profile. This picture well agrees with the numerical evaluation of the free energy $\chi(\beta,w)$ (Fig. \ref{fel}). In particular, for $w\neq 0$ the free energy shows a profile described by (\ref{free0}), while for $w=0$ it dramatically collapses to zero. The geometric morphology of $\chi(\beta,w)$ can be then employed to derive a phase diagram for our thermodynamic system, with the shape of the free energy acting as an order parameter for different observable dynamics. By considering Fig. \ref{fel}, I identify two radically distinct dynamical evolutions:\\
i) for $(\beta,w\neq 0)$ the free energy is of type (\ref{free0}), thereby supporting a first order phase transition in the soliton dynamics. The phase transition, born from the different scaling of the box-counting probability $p_i$, leads to the accumulation of soliton particles with close but opposite eigenvalues (velocities)  towards  gap edges. Such a process, enforced by the action of the parameter $w$, is the signature of a strongly-discontinuous dynamics with solitons expected to immediately escape from their initial positions.\\
ii) for $(\beta,w=0)$, the free energy $\chi$ collapses towards a flat profile $\chi=0$. In this case no phase transitions are predicted, and no cooperative dynamics occurs. When $w=0$, in fact, the input profile $\psi_0$ turns into the trivial constant solution of the NLS $\psi(x,t)=\rho$, which does not show any significant evolution.\\
In order to verify this analysis, I investigate the time domain dynamics of Eq. (\ref{nlsd0}). Figures \ref{ss1}-\ref{ss2} summarize the results for $\rho=30$ and $w$ varying in the range $w\in[0.9,1]$. In agreement with the phase diagram resulting from (\ref{fel}), the system develops a discontinuous type of evolution known as dispersive shock \cite{we_ub_gurev,w2_ub_gurev,mt_op_kamcha,nls_ub_kamcha,sw_ds_hoefe}: the field initially collapses to a singular (wave breaking) point $t^*$ (Fig. \ref{ss1} insets) and then generates a series of filaments escaping from $t^*$ (Fig. \ref{ss2}). As discussed above, such dynamics stems from the many-body solitons cooperation resulting from the first order phase transition supported by $\chi$. The phenomenon of eigenvalue compression, sustained by $w$, is then evident by comparing Figs. \ref{ss1}a-d. This effect can be then employed to control the soliton cooperative dynamics (Fig. \ref{ss1}). As seen in Fig. \ref{ss1}, in fact, a change in $w$ is accompanied by a significant shift of the shock point $t^*$, which increases in time as soon as the grayness is reduced. The positive sign of the shift can be predicted from the free energy collapse at $w\rightarrow 0$. In this condition, $\psi(x,t)=\rho$ and $t^*\rightarrow\infty$, which anticipates an increasing of $t^*$ for a decreasing value of $w$.\\
In summary, the thermodynamics analysis based on the soliton invariant measure predicts the emergence of a first-order phase transition, which gives rise to dispersive shocks with completely controllable dynamics. Such a complex interaction, which cannot be captured with a reductionist analysis focused on the individual system component (i.e., the single soliton), is an emergent property of the whole soliton ensemble.
\section{Solitons and other complex systems}
\label{sec:cm}
As we have seen in the previous section, a many-body ensemble of solitons can act like a complex system, supporting the generation of emergent properties when cooperative dynamics among the soliton particles settles in. In this section, I will compare this system to various dynamics manifesting complex phenomena. This analysis is not only important to highlight unique features of solitons, but also to identify common denominators at the origin of complex behaviors. In the following I will consider two main complexity classes, which also represent traditional examples in statistical mechanics: glassiness and anomalous scaling in time series dynamics.\\
Glassiness is a complex phenomenon whose representative examples, other than in real glasses, are found in a large number of contexts including error correcting codes, supercooled liquids, polymers and neural networks \cite{SGTAB,goetze91:_liquid_freez_and_glass_trans,caangel,sourlas89:_spin_glass_model_as_error_correc_codes}. Although the physics of the glass state is still under active research, different approaches based on mode coupling theory and/or replica-based mean field analysis highlight the same fundamental physics, even from different perspectives. Glassy systems, in particular, show a high degree of frustration \cite{frusta} in their ground state, which appears extremely degenerate and characterized by an incredibly-large number of metastable states of equivalent energy. When the temperature is cooled down below a specific threshold, the complexity of the ground state manifests in the so called glass phase, characterized by an exponential number of energy minima separated by high potential barriers which localize the dynamics in the configuration space. Due to the huge number of energy traps, the system in the glass state explores only a subset of the available phase space. For this reason, glassiness could not be described within the ergodic hypothesis ---which is at the basis of classical statistical mechanics--- and needs a different approach. Breaking of classical thermodynamic assumptions is also observed in the statistical analysis of time series generated by complex dynamics, such as DNA sequences, earthquakes, solar flares and eye movements during spoken conversations \cite{PhysRevLett.84.1832,PhysRevE.52.5281, PhysRevE.66.031906, PhysRevLett.92.138501, PhysRevLett.90.248701,PhysRevE.79.056114}. In all these systems, time sequences are first converted into many distinct trajectories and then into probability distributions. At equilibrium, the underlying system complexity yields probability functions with anomalous scalings, which cannot be interpreted with classical statistical mechanics arguments and whose origin is still largely debated \cite{PhysRevE.79.056114}.\\
In analogy with these systems, solitons complex dynamics are here observed due to the breaking of conventional thermodynamics rules. For the soliton systems, however, breaking is not driven by frustration ---such as in glasses--- but conversely by the lack of any entropy constraint in the soliton invariant measure. The application of a standard canonical measure on a purely kinetic system like solitons, in fact, is a well known results of classical thermodynamics and yields a trivial free energy profile with no cooperative dynamics \cite{SMAST}. The lack of a canonical measure, or equivalently, the absence of any entropy constraint in the soliton invariant measure, is then the key for the appearance of phase transitions and soliton cooperation effects. Such a ``canonical breaking'' is a peculiar property of solitons: in glassy systems, for example, the dynamics within a single energy trap is always described by employing a reduced ergodic assumption, with canonical probabilities satisfying the entropy maximization principle (see e.g., \cite{SGTAB} for more details). Moreover, due to the complete factorization of the Hamiltonian into single-soliton contributions [see Eq. (\ref{nint0})], no frustration exist in the soliton ensemble which exhibits a specific crystalline ground state with all eigenvalues concentrated in a region of the phase as small as possible, where the single-soliton Hamiltonian gets its minimum.\\  
In summary, the breaking of some assumptions of classical statistical mechanics unifies different complexity classes, including solitons, glasses and complex time series. Among them, a distinctive features of solitons is to posses an invariant measure which does not result from any entropy maximization principle.

\section{Conclusions}
I addressed the problem of finding a thermodynamic description of a microscopic mechanical system described by an invariant measure which is not constrained by any a-priori condition. By taking a soliton ensemble as a relevant, interesting case, I demonstrated how to construct an invariant measure from the spectral transform of the system, and how to develop a thermodynamics by employing Lie transformation groups. I have discussed a specific example with a universal model of wave propagation, namely the NLS equation, with reference to a transparent potential supporting an ensemble of gray solitons. For this system I showed, through detail calculations verified by a time-domain analysis, the existence of emergent properties which can be completely controlled by adjusting solitons input parameters. By a further comparison with different complex systems, I have discussed similarities and peculiarities of soliton waves. In particular, like glasses and complex time series, solitons break some assumptions of classical statistical mechanics and this is at the basis of the manifestation of their emergent properties. In the case solitons, in particular, breaking occurs because of the lack of any entropy maximization condition in their invariant measure, and this is a peculiar property of this ensemble. This work is expected to stimulate further theoretical and experimental work aimed at discovering new emergent properties in the broad field of complex many-body dynamics.\\

\begin{acknowledgments}
The research leading to these results has received funding by Award No. KUK-F1-024-21 (2009/2012) made by King Abdullah University of Science and Technology (KAUST). The author thanks S. Trillo for fruitful discussions.\\
\end{acknowledgments}

\appendix

\section{Calculation of the spectral coefficients $c_n$}
\label{cn0}
In the numerator of $c_n$, for $\lambda_{\pm n}$, I expand $b(\lambda_{\pm n})$ in terms of trigonometric functions:
\begin{align}
\label{a1}
  &b(\lambda_{\pm n})=\frac{\sin\pi\rho}{i\sin\pi(\rho-n)}=\frac{(-1)^n}{i}, &n=0,\dotsc,M.
\end{align}
I then use de l'Hopital rule:
\begin{align}
\label{a2}
  \frac{1}{\dot{a}(s_j)}=\lim_{s\rightarrow s_j}\frac{s-s_j}{a(s)},
\end{align}
and then expand the spectral coefficient $a$ in terms of complex fractions:
\begin{align}
\label{a3}
  a(s)=i\prod_{\pm n}\frac{s_j-s_{\pm n}}{s_j-s_{\pm n}^*},
\end{align}
with $s_{\pm n}=\lambda_{\pm n}+2i(\rho-n)$ and $n=0,\dotsc,M$. By substituting Eq. (\ref{a3}) into (\ref{a2}), I obtain:
\begin{align}
\label{a4}
  \frac{1}{\dot{a}(s_j)}=\frac{\prod_{\pm n}(s_j-s_{\pm n}^*)}{i\prod_{\pm n\neq j}(s_j-s_{\pm n})}.
\end{align}
Equation (\ref{a1}), together (\ref{a4}), yields the searched expression of $c_n$:
\begin{align}
  &c_{\pm n}=\frac{b(\lambda_{\pm n})}{\dot{a}(s_j)}=(-1)^{n+1}\frac{\prod_j(s_{\pm n}-s_j^*)}{\prod_{j\neq \pm n}(s_{\pm n}-s_j)}.
\end{align}

%

\end{document}